\documentclass[secnumarabic,amssymb, amsmath,12pt,
nofootinbib,tightenlines, nobibnotes, aps, prl]{revtex4}
\usepackage{graphicx}
\begin{document}
\newcommand{\be}{\begin{equation}}
\newcommand{\ee}{\end{equation}}
\newcommand{\ba}{\begin{eqnarray}}
\newcommand{\ea}{\end{eqnarray}}
\newcommand{\ts}{\textstyle}

\bigskip
\vspace{2cm}
\title{The Unit of Electric Charge and the Mass Hierarchy of Heavy 
Particles}
\vskip 6ex
\author{G. L\'opez Castro}
\email{glopez@fis.cinvestav.mx}
\affiliation{Departamento de F\'{\i}sica, Cinvestav,
Apartado Postal 14-740, 07000 M\'exico, D.F., M\'exico}
\author{J. Pestieau}
\email{pestieau@fyma.ucl.ac.be}
\affiliation{Institut de Physique Th\'eorique, Universit\'e catholique de 
Louvain, \\ Chemin du Cyclotron 2, B-1348 Louvain-la-Neuve, Belgium}
\bigskip
\begin{abstract}
   \hspace*{5mm} We propose some empirical formulae relating the masses 
of the heaviest particles in the standard model (the $W, Z,H$ bosons and 
the $t$ quark)  to the charge of the positron $e$ and the Higgs condensate 
$v$. The relations for the masses of gauge bosons $m_{W} = 
(1+e)v/4$ and $m_{Z}=\sqrt{(1+e^2)/2}\cdot (v/2)$ are in excellent 
agreement with experimental values. By requiring the electroweak 
standard model to be free from quadratic  divergencies at the one-loop 
level, we  find: $m_t=v/\sqrt{2}$ and 
$m_H=v/\sqrt{2e}$, or the very simple ratio $(m_t/m_H)^2=e$.   
\end{abstract}

\maketitle
\bigskip

\section{1. Introduction}

  The discovery of successful empirical relations between the parameters 
of  a given 
physical theory have played an important role in the settlement of more 
general theoretical frameworks to describe such phenomena. Well known 
examples of this were the central role played by the Balmer's series
and Planck's formula in the development of quantum mechanics. In the  
framework of the more general theory, such empirical 
relations can emerge 
in a natural way.

  The parameters of the Higgs potential and the gauge and Yukawa couplings 
are unrelated fundamental parameters in the standard electroweak model 
(SM) of 
particle  physics. They give rise to the observable masses and couplings 
of  particles which are accessible to and can be fixed by experiments. 
Finding 
 successful empirical relations among them may  become useful guides in 
searching the principles behind a more fundamental theory. 

The purpose of this letter is to point out interesting relations which 
seem to relate the unit of electric charge and the masses of the heaviest 
particles in the SM, namely the massive gauge bosons, the top quark and  
the Higgs boson.  We consider that the standard electroweak 
model is an effective model embedded within a more fundamental 
theory and that results obtained at the one-loop level in the SM gives a 
good approximation to the real world. Our proposal assumes that the unit 
of electric charge $e$ and the vacuum expectation value of the Higgs 
field $v$ play a fundamental role in such general theory. We mean by this 
that the mixing  angle of 
neutral gauge bosons $\theta_W$ and the masses of such heavy particles can 
be  expressed solely in terms of $e$ and $v$ . Thus, we propose empirical 
relations 
relating these parameters to the observable physical masses of heavy 
particles which turn out to be in surprisingly  good agreement with 
experiment.

\section{2. Masses of gauge bosons}

 We start with the electroweak standard model \cite{1,2,3} which is based 
on the gauge group $SU(2)_L\otimes U_Y(1)$, with associated gauge boson 
fields  $W^{i}_{\mu}\ (i=1,2,3)$ and $B_{\mu}$. After spontaneous symmetry 
breaking, $W^{i}_{\mu}$ and $B_{\mu}$ acquire masses $m_{W}$ and $m_{B}$,  
respectively, and $W^3_{\mu}$ and $B_{\mu}$ get 
mixed. The diagonalization of the mass matrix for the neutral components, 
gives rise to the physical fields $A_{\mu}$ and $Z_{\mu}$ corresponding to 
the massless photon and the neutral $Z$ boson of mass $m_Z$, 
satisfying the following relations 
\cite{2,4},
 \begin{eqnarray}
m^2_{Z} 
&=& m^2_{W} +m^2_{B}
\label{1}\\
\cos \theta_{W} 
&=& \frac{m_W}{m_{Z}} \ \ \ , \ \ \ \sin
\theta_{W} = \frac{m_B}{m_{Z}}\ , \label{2}
\end{eqnarray}
where $\theta_{W}$ is the weak mixing angle:
\begin{eqnarray}
Z_\mu &=& W^3_\mu \cos \theta_{W} - B_\mu \sin
\theta_{W}\ , \label{3}\\
A_\mu &=& W^3_\mu \sin \theta_{W} + B_\mu \cos \theta_{W}\ . \label{4}
\end{eqnarray}

 We propose two remarkable empirical mass  relations defining two 
different mass scales: 
\begin{eqnarray}
m_{W} + m_{B} &=& \frac{v}{2} \ ,\label{5}\\
 m_{W} - m_{B}&=& e \frac{v}{2} \ , \label{6}
\end{eqnarray}
where $e$ is the positron electric charge and $v$, the strength of the Higgs
condensate. From the above equations it is easy to derive the following 
relations \cite{5,6}:
\begin{equation}
e = \frac{1-\tan \theta_{W}}{1 + \tan \theta_{W}} = \tan \left( 
\frac{\pi}{4} -\theta_{W}\right)\ , \label{7}
\end{equation}
and
\begin{equation}
m^2_{Z} = (1+e^2) \cdot\frac{v^2}{ 8}\ . \label{7'}
\end{equation}

Now, using as experimental inputs the values of the fine structure 
constant and the $Z$ boson mass \cite{7}, $\alpha = e^2/4\pi = 
(137.03599911(46))^{-1}$ and $m_{Z} = 91.1876(21)$ GeV we can fix $e$ and 
$v$, thus we obtain:
\begin{eqnarray}
\tan \theta_{W} &=& \frac{1-e}{1+e} = 0.53513    
      \label{10}\\
v &=& 246.8476 \pm 0.0057 \ 
\textrm{GeV}                                         
\label{11}\\
m_{W} &=& \frac{v}{ 4} (1+e) = 80.400 \pm 0.002 \
\textrm{GeV}\ ,  \label{12}\\
m_{B} &=& \frac{v}{ 4} (1-e) = 43.024 \pm 0.001 \
\textrm{GeV} \ ,   \label{13}
\end{eqnarray} 
 to be compared with the experimental values \cite{7}:
 \begin{eqnarray}
\tan \theta_W (\exp) &=& \sqrt{m_Z^2/m_W^2-1}=0.53503 \pm 0.00087 \ , \\ 
m_{W} (\exp) &=& 80.403 \pm 0.029 \ \textrm{GeV}           
           \label{14}\\
   v_{F} &\equiv & \left( 
\sqrt{2}G_{F}\right)^{-1/2} = 246.221 \pm 0.001\
\textrm{GeV}  \ ,                    \label{15}
\end{eqnarray}
where $v_{\textrm{\tiny\textit{F}}}$ is the usual value of  the Higgs
condensate defined from the Fermi constant $G_{F}$.  The 
agreement between the predictions based on our proposed relations, Eqs. 
(\ref{5}) and (\ref{6}), and experimental values is impressive. The 
largest difference is found in the value of $v$ although it differs 
from $v_F$ only at the per mille level: $v/v_F -1 = 0.0025$.
Thus, we can conclude that Eqs (\ref{5}) and (\ref{6}) are robust 
relations and may  point to new physics.

 Before we close this section, let us speculate a bit about the 
possible origin 
of the mass relations proposed in Eqs. (5,6). Let us imagine an scenario 
where the vector gauge fields $W^i_{\mu}$ and $B_{\mu}$ arise from the 
direct product of even more fundamental isodoublet fields (in analogy to 
strong interactions of ($u,d$) quark fields giving rise to the triplet 
$\rho_i^{\mu}$ and singlet $\omega^\mu$ meson fields). Let us call 
$T_{\mu}$ and $S_{\mu}$ the neutral  components of this  direct product, 
such that: 
\begin{eqnarray}
T_\mu &=& \frac{1}{\sqrt{2}} (W^3_\mu -B_\mu),  \label{17}\\
S_\mu &=& \frac{1}{\sqrt{2}} (W^3_\mu +B_\mu). \label{18}
\end{eqnarray}
A possible hierarchical mass pattern for the system of ($T,S$) neutral 
vector fields and its expression in terms of SU(2) isotriplet and 
isosinglet fields are:
\ba
-2 {\cal L}_{M} &=& \left( T_\mu,\ S_\mu \right)
\frac{v^2}{8} \left(\begin{array}{ccc} 1 & e \\ e & e^2
\end{array}\right) \left(\begin{array}{c} T_\mu \\ S_\mu
\end{array}\right) \ \nonumber \\
&=& \left( W^3_\mu,\ B_\mu \right)
\left(\begin{array}{lll} \ \ m^2_{W} 
& -m_{W} m_{B} \\ 
-m_{W} m_{B} 
& \ \  m^2_{B}\end{array}\right)
\left(
\begin{array}{c}W^3_\mu
\\ B_\mu \end{array}\right)\nonumber \\
&=& \left(Z_{\mu},\ A_{\mu} \right) 
\left( \begin{array}{ll} \ \ m_Z^2 & 0 \\ 0 & 0 \end{array}\right)
\left(\begin{array}{c} Z_{\mu} \\ A_{\mu} \end{array}\right)\ 
. \label{21}
\ea
This hierarchical pattern of masses for $T_{\mu}$ and $S_{\mu}$ fields 
gives the following mass relations:
\ba
m_{T} &=&  \frac{1}{\sqrt{2}} (m_{W}+ m_{B})=\frac{v}{2\sqrt{2}} \ , 
\label{19}\\
m_{S} &=& \frac{1}{\sqrt{2}} (m_{W} - m_{B}) = \frac{ev}{2\sqrt{2}}\ . 
\label{20}
\ea
Thus, these hierarchical mass relations for the $T,S$ vector fields would 
naturally reproduce our proposed relations given in Eqs. (5,6). In 
particular, the unit of charge $e$ plays a central role in determining 
this hierarchy since $m_S/m_T=e$.

\section{3. Fixing the value of $e$}

  As we have seen, the charge of the positron $e$ plays a central  
role in our empirical relations for the masses of gauge bosons. In this 
section we explore a very simple formula which allows to fix the value of 
$e$ and will simplify our expressions for the top quark and Higgs 
boson masses to be discussed below. For this 
purpose, let us start from the formula relating the mass of 
the $W$ boson to the parameters  $e,G_{F}$ and $\sin
\theta_{W}$ \cite{7}:
\begin{equation}
m_W = \frac{e}{ \sin \theta_W}
\frac{1}{(1-\Delta r)^{1/2}} 
\frac{v_F}{2}
\label{26}
\end{equation}
where $\Delta r$ includes the effects of radiative corrections.  

Now, if we compare Eqs. (11) and (21), using first Eq. (9), we find
\begin{equation}
m_{W} = (1+e)\frac{v}{ 4} = \frac{ex}{ 1-e} \cdot \frac{v}{ 4}\ ,
\label{28}
\end{equation}
with \footnote{We can invert 
this formula to obtain a value for the size of radiative corrections 
in terms of $e,v$ and $v_F$. The value obtained $\Delta r=0.03416$ is 
consistent (within less than 2$\sigma$'s) with the theoretical value 
computed in the electroweak SM without cut-off \cite{7}, namely $\Delta 
r=0.03630\mp 0.0011 \pm 0.00014$.} 
\begin{equation}
x= \frac{1}{e}-e= 2\sqrt{\frac{2(1+e^2)}{(1-\Delta r)}}\cdot 
\frac{v_F}{ v}.
\label{29}
\end{equation}
On the other hand, using the value of $\alpha$ given in section 2 we get 
\begin{equation}
x = \frac{1}{ e} - e = 2.99944654\ ,
\label{30}
\end{equation}
which to a very good approximation we can parametrize as
\begin{equation}
\frac{1}{ e} - e = 3 \left(1 - \frac{\alpha}{4\pi^2} + \cdots \right).
\label{31}
\end{equation}
This suggests that, in a not yet known fundamental theory, the positron
electric charge $e$ would be fixed by a very simple relation. In other 
words we propose that, before radiative corrections, the ``bare" finite 
electric charge $\bar e$  should satisfy the following simple relation 
\cite{5}:
\begin{equation}
\frac{1}{\bar{e}} - \bar e = 3.
\label{32}
\end{equation}

Using, in very good approximation, this simple relation we can rewrite 
Eqs. (8) and (11) as follows:
\ba
m_W &=& \frac{3e}{1-e}\cdot \frac{v}{4}\ , \nonumber \\
m_Z &=& \frac{m_W}{\cos \theta_W} = \sqrt{2(1+e^2)}\cdot \frac{3e}{1-e^2} 
\frac{v}{4}\ . \nonumber 
\ea
   The above expressions are reminiscent of the expected dependence 
of the gauge  boson masses in the SM, namely that they are of first order 
in the gauge  coupling. 

\section{4. The  top quark and the Higgs boson masses}

In order to find our relations for  the masses of the Higgs boson and 
the top quark, we  consider our expressions (\ref{7'}) and (\ref{12}) for 
the masses of the gauge bosons and we use Eq. (\ref{31}) in first 
approximation, namely $1/e-e=3$. We find:
\begin{equation}
2m^2_W + m^2_Z = \left(4 -\frac{1}{e}\right)\cdot \frac{v^2}{2} \ .
\label{34}
\end{equation}

Now, if we compare  Eq (27) with the squared mass sum rule  
\cite{9,10}
\begin{equation}
2m^2_W + m^2_Z = 4m_t^2-m^2_H
\label{36}
\end{equation}
which is the condition to cancel, at the one-loop level, the 
quadratic  divergences  in the standard electroweak theory\footnote{Note 
that if additional heavy degrees of freedom are present like in 
some extensions of the SM, they will modify Eq. \ref{36}.}, 
the following results are naturally suggested\footnote{More general 
expressions that satisfy Eqs. (\ref{34},\ref{36}) are: 
$m_t^2=[2+X](v^2/4)$ and $m_H^2=[1/e+2X](v^2/2)$, where $X$ is an 
arbitrary but small parameter (Eqs. (29,30) correspond to the simplest 
choice $X=0$). Another interesting choice is $X=-(1+e)^2/24$. This 
value corresponds to the simultaneous cancellation of quadratic {\it 
 and} logarithmic divergencies in the self-mass of the top quark in 
the SM \cite{nos}. The corresponding values of the heaviest 
particles in this case become $m_t=171.43$ GeV (which perfectly matches 
the most recent result $m_t = (171.4 \pm 2.1)\ \mbox{GeV}$ reported in 
\cite{8}) and $m_H=310.35$ GeV.} 
for the masses  of the top quark and the Higgs boson: 
\ba 
m_t&=&\frac{v}{\sqrt{2}}= 174.5\  \mbox{GeV} \ , \label{37} \\ m_H &=& 
\frac{v}{\sqrt{2e}}= 317.2\ \mbox{GeV}\ , \label{38} 
\ea
with the very interesting ratio $(m_t/m_H)^2=e$.

The mass of the top quark found in Eq. (29) is in excellent agreement 
with the direct measurements obtained at $p\bar{p}$ 
colliders [$m_t=(174.2\pm 3.3)$ GeV]  \cite{7} and from fits to 
electroweak  data [$m_t=(172.3^{+10.2}_{-7.6})$ GeV] \cite{7}.  On the 
other hand, the mass of  the Higgs boson shown in Eq. (30) is almost 
completely ruled out by  
the bounds obtained from fits to electroweak data in the SM \cite{7}:
$m_H < 186 \ \textrm{GeV\ at \ 95\% \ c. l.}$.

Nevertheless, if the renormalizable SM is embedded into a more fundamental 
renormalizable theory (in the same way as QED in embedded into the 
Electroweak SM), the effects of the new physics scale $\Lambda_{NP}$  can 
be felt in physical observables at the electroweak scale. For instance, if 
the  virtual effects of  $\Lambda_{NP}$ in the determination of the $W$ 
boson  mass are  of 
$O(0.1\%)$ this could compensate the effects of a heavier Higgs boson 
\cite{11,12}. In this case, a heavier SM Higgs boson of $300\sim 320$ GeV 
could be perfectly accommodated by present electroweak data and can be 
produced and observed by the LHC experiments.  

\section{5. Conclusions}

Summarizing, we have proposed a set of simple empirical formulae for 
the masses of gauge bosons in the effective Standard Electroweak Model, 
Eqs. (5) and (6), 
which turns out to be in excellent numerical agreement with present data. 
Further simple formulas  are derived for the top quark and Higgs boson 
masses when we add the requirement of cancellation of quadratic 
divergencies (at the one-loop level) in the self-masses. In particular, 
the mass of the Higgs boson is predicted to lie in the range 300$\sim$320 
GeV which can be accessible at the LHC collider. On the other 
hand, although present electroweak data seems to rule out this heavy 
Higgs boson, sufficiently small effects of new physics can relax the upper 
bound that is allowed by present data. 

    In such new physics scenario, the unit of electric charge $e$ should 
play an essential role as a fundamental parameter. We have proposed also a 
very simple and elegant formula which would fix the value of $e$ in the 
new physics theory. The simplicity and symmetry of equation (26) (note 
that it remains invariant under the transformation $\bar{e} 
\leftrightarrow -1/\bar{e}$ ) suggest that a  dual symmetry
may be a must for the underlying theory. 

   Finally, we would like to emphasize that our formulae for the masses of 
heavy particles are satisfied by the physical masses and not by running 
masses defined at an arbitrary scale. As another example of 
this, let us remember the sum rule involving the masses of charged leptons 
proposed in ref. \cite{koide}: 
$\sum_l m_l = (2/3)[\sum_l\sqrt{m_l}]^2$. This sum rule predicts the mass 
of the heaviest lepton to be $m_{\tau}=1776.969$ GeV, which perfectly 
matches the value of the physical mass measured by $e^+e^-$ colliders at 
the $\tau$-pair production threshold \cite{7}, but fails to 
be satisfied by the running mass values.

\section{Acknowledgements}
We are pleased to express our gratitude for the useful comments received 
from Jan Govaerts, Mat\'\i as Moreno and Christopher Smith. The work of GCL
has been partially supported by Conacyt.


\begin{thebibliography}{90}
\bibitem{1} S.L. Glashow, {\it Nucl. Phys.} {\bf 22} (1961) 579.
\bibitem{2} S. Weinberg, {\it Phys. Rev. Lett.} {\bf 19} (1967)
1264.
\bibitem{3} A. Salam, in {\em Elementary Particle Physics:
Relativistic Groups and Analyticity}, Ed. N. Svartholm (Almqvist
and Wiksell, 1968).
\bibitem{4} J. Pestieau and P. Roy, {\it Phys. Rev. Lett.} {\bf 23}
(1969) 349; {\it Lett. Nuovo Cimento} {\bf 31} (1981) 625.
\bibitem{5} G. Lopez-Castro and J. Pestieau, e-print archive 
hep-ph/9804272.
\bibitem{6} J. Pestieau, e-print archive hep-ph/0105301.
\bibitem{7} W.-M. Yao et al {\it Review of Particle Physics}, J.of Phys. 
{\bf G33}, (2006) 1.
\bibitem{9} R. Decker and J. Pestieau, UCL-IPT-79, Presented at DESY Workshop,
October 1979; reprinted in hep-ph/0512126.
\bibitem{10} M. Veltman, \textit{Acta. Phys. Pol.} \textbf{B12} (1981) 437.
\bibitem{nos} G. L\'opez Castro and J. Pestieau, Mod. Phys. Lett. {\bf 
A10}, 1155 (1995).
\bibitem{8} E. Brubaker et al (The Tevatron Electroweak Working Group), 
eprint Archive hep-ex/0608032.
\bibitem{11} J. Erler, eprint Archive hep-ph/0604035.
\bibitem{12} J. A. Casas, J. R. Espinosa and I. Hidalgo, eprint Archive 
hep-ph/0607279.
\bibitem{koide} Y. Koide, Lett. Nuovo Cimento {\bf 34}, 201 (1982);
 Phys. Rev. {\bf D28}, 252 (1983); e-print archive hep-ph/0506247.
\end{thebibliography}
\end{document}